\documentclass[twocolumn]{aastex63}
\usepackage{mhchem}

\submitjournal{ApJ}
\shorttitle{Type-IIP Supernovae}
\shortauthors{Jiang et al.}

\begin{document}

\title{A Model for Type-IIP Supernovae with Medium Recombination Approximation}
\correspondingauthor{Xue-Wen Liu}
\email{liuxuew@scu.edu.cn}

\author{Hong-Xuan Jiang}
\thanks{hongxuan\_jiang@sjtu.edu.cn}
\affiliation{Tsung-Dao Lee Institute, Shanghai Jiao Tong University, Shengrong Road 520, Shanghai, 201210, People's Republic of China}
\affiliation{College of Physics, 
Sichuan University, Chengdu, 610065, People's Republic of China}

\author{Xue-Wen Liu}
\affiliation{College of Physics, 
Sichuan University, Chengdu, 610065, People's Republic of China}

\begin{abstract}
In this paper, we propose a new light-curve model for Type-IIP supernovae (SNe IIP) with an approximation of medium recombination. Recombination of hydrogen that takes place in the envelope is believed highly affect the light curves of SNe IIP. The propagation of the recombination wave through the expanding envelope is crucial to determine the temperature and the bolometric luminosity. Several approximations were made in previous works to determine the recombination front, which plays a role as the pseudo-photosphere. 
With the Eddington boundary condition, we let the actual photosphere be the boundary to determine the time evolution of temperature profile of the envelope and calculate the bolometric luminosity. A new approximation on the speed of recombination wave is introduced to get a closer result to the real situation. We abandon the non-self-consistent approximations made in former works and solve the initial hump problem in the previous attempt for the slow recombination approximation. The produced light curves show the necessity of this approximation and fit the observation well. 

\end{abstract}


\section{Introduction} \label{sec:intro}
{Type II and a subset of Type I (Ib/Ic; stripped envelope) supernovae are widely accepted to be originated from the core collapse of stars with zero-age-main-sequence mass of $\geq 8\,M_{\odot}$ \citep{2015PASA...32...16S}. SNe II are defined by the presence of hydrogen in their spectra and classified according to their shape of light curves: Type IIL with a linearly declining light curve, whereas Type IIP displaying an extended plateau lasting $\sim 100$ days. 
Despite some authors find growing evidence that SNe II form a continuum of light-curve properties, the separation between SNe IIP and IIL is still an open issue \citep{Arcavi2012, Faran2014, Anderson2014, Sanders2015, Valenti_2016, Galbany2016, Rubin2016, Hillier2019}. \cite{Kou2020} found that the profile of the $H_\alpha$ spectral line may be the key to distinguishing the two types of SNe II. However, it is still an open question.}

Indeed, the amount and distribution of hydrogen is one of the most important ingredients in shaping the long-lasting optical plateau of SNe II through recombination effect \citep{Grassberg1971, Falk1973, 1989ApJ...340..396A,1993ApJ...414..712P, Kasen2009, Dessart2013,Nagy2014,Faran2019}. The spatial distribution of 
 \ce{^{56}Ni} also plays an important role in extending and flattening the plateau of the light curves \citep{Nakar2016, Kozyreva2019}. One-dimensional (1D) radiation hydrodynamical code \texttt{STELLAR} or radiation transport code \texttt{Sedona} based on 1D progenitor model \texttt{MESA} including the above effects can synthesize the light curves of SNe IIP \citep{Paxton2018, Paxton2019, Goldberg2019, Kasen2009, Ricks2019}. Three-dimensional neutrino-driven explosion simulations can now explain the basic observational data of ordinary SNe IIP \citep{Utrobin2017}. {However, a simple analytical method without complicated, time-consuming simulations is still useful to obtain the basic physical parameters such as explosion energy, ejected mass and initial radius for the first order-approximation. Moreover, the analytical scalings between the parameters have a clear physical explanation and are easily compared with the observations.}

An analytical model for SNe was proposed by Arnett which described the light curve properties of SNe I and SNe II in detail \citep{1980ApJ...237..541A, 1982ApJ...253..785A, 1989ApJ...340..396A}. With the help of Arnett's model, some important parameters, e.g., ejecta mass, explosion energy, and initial radius can be estimated quickly \citep{Nagy2014}. \cite{1989ApJ...340..396A} discussed the effects of ionization on the light curves of SNe II under the approximation of fast recombination. Two approximations about the speed of hydrogen recombination in SNe IIP are summarized in \cite{Arnett1996}, in which slow recombination approximation was introduced. {For the fast recombination model, the temperature profile remains unchanged when the recombination front recedes through the envelope.} While the slow approximation assumes the temperature profile of the envelope evolves {strictly} with the recession of the recombination front \citep{Arnett1996}. Based on the fast recombination approximation, \cite{Nagy2014} successfully produced the light curves of SNe IIP and fitted observations very well. Although their model can not explain the cusp of the light curves of SNe IIP at very early epochs, their model demonstrates the effect of recombination is the critical factor in the formation of the light curves of SNe IIP. In the following paper, they solved the cusp problem by using a two-component model \citep{Nagy2016}. 
 
The main difference between the fast and slow approximations is the boundary condition of the spatial equation of the model. The fast approximation is widely assumed in previous works (see \cite{Nagy2014, 2003MNRAS.338..711Z, Chatzopoulos2012}). The boundary is simply set to be the outmost edge of the hydrogen-rich envelope. However, the slow approximation requires calculating the real-time temperature profile of the envelope. Thus time-dependent boundary is required. In \cite{1980ApJ...237..541A}, the Eddington surface boundary condition was introduced, giving constraints on {the spatial temperature profile} $\psi$ at the photosphere \citep{1926ics..book.....E}. 
Assuming the receding recombination front as the pseudo-photosphere, \cite{1989ApJ...340..396A} implied that the eigenvalue $\alpha$ {of the temperature profile} increases in proportion to $1/{x_{\rm i}}^2$, where $x_i$ is the dimensionless radius of the recombination front. \cite{Arnett1996} implemented this idea to the calculation of light curve. However, the produced light curve has a big bump at the end of the plateau phase, which is not physical. {Using the "radiative-zero" boundary condition \citep{1980ApJ...237..541A}, \cite{1993ApJ...414..712P} assumed the temperature at the recombination front is zero and pointed out the necessity of time-dependent radial temperature profile. His result is also consistent with the relation of $\alpha(t) \propto 1/{x_{\rm i}^2(t)}$. $\alpha$ is the eigenvalue for the spatial equation, which describe the shape of the radial temperature profile. However, the zero surface temperature assumption is too rough to strictly fit observation. More detailed model is required.} 

{Recently, \cite{2018ApJ...868L..24L} considered the recession of the real photosphere and derived its radius evolution, which is consistent with observed supernova-like explosion. \cite{2022ApJ...933....7J} then propose a method to obtain the light curve when the photosphere recession effect is presented.}
In this paper, we further add the recombination effect to construct a semi-analytical light curve model for SNe IIP. We can remove the approximations made in \cite{1989ApJ...340..396A}, which result in the bump during the plateau phase. We adopt a boundary condition on the recession photosphere to determine the instantaneous temperature profile, which leads to the coevolution of $\alpha$ and $x_{\rm ph}$. The SN ejecta is homologous, spherically symmetric and the density profile is exponential or uniform. The opacity is assumed to be a simple step-function of temperature introduced in \cite{1989ApJ...340..396A}. 

This paper is organized as follows. { In Section \ref{Sec: 2}, we describe the model. In Section \ref{Sec: 3}, we describe the medium recombination approximation in detail. In Section \ref{Sec: 4}, we compare our model with previous models. In Section \ref{Sec: 5}, we discuss the effects of each parameter. In Section \ref{Sec: 6}, we describe the shell component in our model. In Section \ref{Sec: 7}, we fit 5 well studied sources. Finally, in Section \ref{Sec: 8}, we give our conclusion. }

\section{Model construction and approximations} \label{Sec: 2}
\subsection{The light curve model} \label{Sec: 2.1}
Based on the model of \cite{1980ApJ...237..541A}, \cite{1989ApJ...340..396A} added the recombination effect. In this section, we modify some of their assumptions and consider the effect of photosphere recession and thus time-dependent boundary. 

Taking the same separation of variables in \cite{1980ApJ...237..541A}, we can separate the temporal and spatial part of temperature as follows
\begin{equation}
	T^4(x, t) = T^4(0, 0)\psi(x)\phi(t)\left(\frac{R_{\rm 0}}{R(t)}\right)^4, \label{Eq: temperature}
\end{equation}
where $\psi(x)$ and $\phi(t)$ are the spatial and temporal profiles, respectively. Since the ejecta is homologous, we define $x$ to be a dimensionless variable which satisfies $x=r/R(t)$. In the same way, the density profile can also be separated as follows
\begin{equation}
	\rho(x, t) = \rho_{\rm 0}\eta(x)\left(\frac{R_{\rm 0}}{R(t)}\right)^3, \label{Eq: density}
\end{equation}
where $\eta(x)$ is the spatial distribution. 

In the diffusion approximation, the luminosity $L$ is
\begin{equation}
	\frac{L}{4\pi r^2} = -\frac{\lambda c}{3}\frac{\partial aT^4}{\partial r}, \label{Eq: diffusion approximation}
\end{equation}
where $\lambda=1/\rho\kappa$ {is the mean free path}, $\kappa$ is opacity, and {$a$ is the Stefan–Boltzmann constant. }

We assume the {extra} energy source is the release of nuclear energy by \ce{^{56}Ni} decay only (which is also the assumption in \cite{1982ApJ...253..785A}). The nickel \ce{^{56}Ni} turns into \ce{^{56}Co} through radioactive decay. The decay of \ce{^{56}Ni} and \ce{^{56}Co} provides the energy source for SNe II and forms a radioactive tail in the light curves \citep{1984A&A...135...27B}. The abundance equations can be written as
\begin{equation}
	\frac{dX_{\rm Ni}(t)}{dt}=-\frac{X_{\rm Ni}(t)}{\tau_{\rm Ni}} \label{Eq: Ni},
\end{equation}
\begin{equation}
	\frac{dX_{\rm Co}(t)}{dt}=\frac{X_{\rm Ni}(t)}{\tau_{\rm Ni}}-\frac{X_{\rm Co}(t)}{\tau_{\rm Co}}\label{Eq: Co},
\end{equation}
where
$\tau_{\rm Ni}$ and $\tau_{\rm Co}$ are the decay time of \ce{^{56}Ni} and \ce{^{56}Co} respectively. As defined in \cite{1982ApJ...253..785A}, the expression of the energy release $\epsilon$ can be written as
\begin{equation}
	\epsilon = \epsilon_{\rm Ni} M_{\rm Ni} \zeta(t), \label{Eq: epsilon}
\end{equation}
where 
\begin{equation}
	\zeta(t) = X_{\rm Ni}(t) + \frac{\epsilon_{\rm Co}}{\epsilon_{\rm Ni}}X_{\rm Co}(t). \label{Eq: zeta}
\end{equation}
$\epsilon_{\rm Ni}$ and $\epsilon_{\rm Co}$ are the energy production rate via radioactive decay.

Note that the first law of thermodynamics can be written as
\begin{equation}
	\frac{dE}{dt}+P\frac{dV}{dt}=\epsilon-\frac{\partial L}{\partial m}, \label{Eq: the first law of thermodynamics},
\end{equation}
where $E$ is thermal energy per unit mass, $P$ is pressure and $V = 1/\rho$. For radiation gas, we have energy density $E=aT^4V$ and pressure $P=aT^4/3$.

The temperature of the hydrogen-rich envelope of SNe II is more than $10^6 \,\rm K$ at the very beginning, which means the whole envelope is fully ionized. As the envelope expands outwards, the temperature cools down gradually. Adopting the same assumption in \cite{1989ApJ...340..396A}, when the temperature is lower than recombination temperature $T_{\rm ion}$, the local opacity is zero. {For the plasma with temperature higher than $T_{\rm ion}$, its opacity is a constant $\kappa_{\rm t}$.} Therefore a recombination front $x_{\rm i}$ divides the envelope into two parts, and the opacity function can be written as
\begin{equation}
\kappa(T)=\left\{
	\begin{array}{lll}
		\kappa_{\rm t},  &  & T>T_{\rm ion}\\
		0,         &  & T<T_{\rm ion}
	\end{array}.
	\right. \label{Eq: kappa}
\end{equation}

As the same in \cite{1980ApJ...237..541A} and \cite{2018ApJ...868L..24L}, 
we also choose optical depth $q=2/3$ to determine the photosphere $x_{\rm ph}$. This value comes from the approximation of a plane-parallel atmosphere. Using the equation in \cite{2018ApJ...868L..24L}
\begin{equation}
	\int_{r_{\rm ph}}^{r_{\rm i}} \rho(x, t) \kappa_{\rm t} dr=2/3, \label{Eq: optical depth}
\end{equation}   
where {$r_{\rm i}$ is the radius of the recombination front.} {For the rest of calculation, we} assume an exponential density distribution
\begin{equation}
	\eta(x)=e^{Ax}. \label{Eq: eta}
\end{equation} 
Substituting $\rho(x, t)$ into Eq.~(\ref{Eq: optical depth}), we obtain the expression of $x_{\rm ph}$ \citep{2022ApJ...933....7J}:
\begin{equation}
	x_{\rm ph} = \frac{1}{A}\ln\left[e^{A x_{\rm i}}-A \frac{2R^2(t)}{3R_0^3 \kappa_{\rm t}}\right],\label{Eq: xph}
\end{equation}
{where $x_{\rm i}$ is dimensionless radius of the recombination front, i.e. $T(x_{\rm i})=T_{\rm ion}$.}

Photons that reach the photosphere can escape freely. Therefore the contribution outside of the photosphere is neglected. Using the first law of thermodynamics, we have
\begin{equation}
	\dot E_{\rm ph}+\left(P\dot V\right)_{\rm ph} = \epsilon_{\rm ph} - L_{\rm ph}. \label{Eq: int to photosphere}
\end{equation}
Define
\begin{equation}
\begin{aligned}
E_{\rm ph} = &\int_0^{r_{\rm ph}}a T^4 4\pi r^2dx \\
= & E_{\rm Th}^0\frac{I_{\rm ph}}{I_{\rm Th}^0}\frac{R_{\rm 0}}{R(t)}\phi(t), \label{Eq: Eph}
\end{aligned}
\end{equation}
where $E_{\rm Th}^0 = 4\pi R_0^3 a T_0^4 I_{\rm Th}^0$ {is the initial thermal energy of the envelope}, $I_{\rm ph}=\int_0^{x_{\rm ph}}\psi(x)x^2dx$ and $I_{\rm Th}^0=\int_0^1\psi^0(x)x^2dx$ are the integration of the radial profile. $\psi^0(x)$ is the initial radial temperature profile, while $\psi(x)$ varies with the recession of photosphere.
Thus the time derivative of it is written as
\begin{equation}
\begin{aligned}
	\dot E_{\rm ph} = &E_{\rm ph}\left(\frac{\dot \phi}{\phi}-\frac{v_{\rm sc}}{R(t)} + \frac{\dot I_{\rm ph}}{I_{\rm ph}} \right)\\
	 = &E_{\rm ph}\left(\frac{d\ln{\phi}}{dt}-\frac{d\ln R}{dt} + \frac{d\ln{I_{\rm ph}}}{dt} \right). \label{Eq: left side}
\end{aligned}
\end{equation}
Note that as in \cite{1989ApJ...340..396A}:
\begin{equation}
	\frac{d\ln V}{dt} = 3\frac{d\ln R}{dt},
\end{equation}
and pressure $P=aT^4/3$, we rearrange Eq.~(\ref{Eq: left side}) as follows
\begin{equation}
	\dot E_{\rm ph} + E_{\rm ph}\frac{d\ln R}{dt}=E_{\rm ph}\left(\frac{\dot \phi}{\phi}+ \frac{\dot I_{\rm ph}}{I_{\rm ph}} \right), \label{Eq: left1}
\end{equation}
which is exactly the left side of Eq.~(\ref{Eq: int to photosphere}). Substitute Eq.~(\ref{Eq: left1}) into the L.H.S of Eq.~(\ref{Eq: int to photosphere}), we obtain
\begin{equation}
	\dot E_{\rm ph}+\left(P\dot V\right)_{\rm ph} = E_{\rm ph}\left(\frac{\dot \phi}{\phi}+ \frac{\dot I_{\rm ph}}{I_{\rm ph}} \right), \label{Eq: left}
\end{equation}
where the time derivative of $I_{\rm ph}$ is written as
\begin{equation}
	\dot I_{\rm ph} = \psi(x_{\rm ph})x_{\rm ph}^2\dot x_{\rm ph}
\end{equation}

Substituting Eq.~(\ref{Eq: temperature}) into Eq.~(\ref{Eq: diffusion approximation}) we obtain:
\begin{equation}
\begin{aligned}
	L(x, t) = & - \frac{4\pi a T^4(0,0) c R_0}{3 \rho_0 \kappa_{\rm t}} \phi(t)\frac{x^2}{\eta(x)}\frac{d\psi(x)}{dx}\\
	=& -\frac{E_{\rm Th}^0}{\tau_{\rm d} I_{\rm Th}^0 \alpha}\phi(t) \frac{x^2}{\eta(x)}\frac{d\psi(x)}{dx},
\end{aligned} \label{Eq: luminosity}
\end{equation}
where $$\tau_{\rm d} = \frac{3\rho_0\kappa_{\rm t} R_0^2}{\alpha c}
$$
Thus combining luminosity term Eq.~(\ref{Eq: luminosity}), radioactive decay term Eq.~(\ref{Eq: epsilon}) and Eq.~(\ref{Eq: left}), we can obtain a partial differential equation. Using the widely used technique in \citep{1980ApJ...237..541A, 1982ApJ...253..785A, 1989ApJ...340..396A}, this equation can be separated into temporal and spatial parts, which are shown as follows.

Spatial part:
\begin{equation}
	\frac{1}{\psi(x)x^2}\frac{d}{d x}\left(\frac{x^2}{\eta(x)}\frac{d\psi}{dx}\right) = -\alpha , \label{Eq: spatial part}
\end{equation}
where $\alpha$ is the eigenvalue. As our previous work (see \cite{2022ApJ...933....7J}) suggests, $\alpha$ varies with the recession of photosphere.

Substituting the definition of $I_{\rm ph}$ into Eq: (\ref{Eq: spatial part}), the complicated spatial term in Eq.~(\ref{Eq: luminosity}) can be replaced with a simple $I_{\rm ph}$. 
\begin{equation}
\begin{aligned}
	\frac{x^2}{\eta(x)}\frac{d\psi}{dx} = &-\alpha \int \psi(x)x^2dx\\
	=&-\alpha I_{\rm ph}. \label{Eq: substitution}
\end{aligned}
\end{equation}
And the luminosity can be simplified as
\begin{equation}
	L(x, t) = \frac{E_{\rm Th}^0}{\tau_{\rm d}}\frac{I_{\rm ph}}{I_{\rm Th}^0} \phi(t) \label{Eq: luminosity1}.
\end{equation}
After some algebra, the temporal part is obtained
\begin{equation}
	\frac{d \phi(t)}{dt} = \frac{R(t)}{R_0}\left[p_1 \frac{I_{\rm Th}^0}{I_{\rm ph}}\zeta(t)-\frac{\phi(t)}{\tau_{\rm d}}\right]-\phi(t)\frac{\psi(x_{\rm ph})x_{\rm ph}^2}{I_{\rm ph}}\frac{dx_{\rm ph}}{dt} ,\label{Eq: phi}
\end{equation}
where $p_1 = \epsilon_{\rm Ni}M_{\rm Ni}^0/E_{\rm Th}^0$.

Finally, the effects of gamma-ray leakage and recombination heat from Hydrogen can be added to the luminosity as
\begin{equation}
	L(t) = \frac{E_{\rm Th}^0}{\tau_{\rm d}}\frac{I_{\rm ph}}{I_{\rm Th}^0} \phi(t) \left(1-e^{A_{\rm g}/t^2}\right) + 4 \pi^2 r_{\rm i}^2 Q\rho \left(x_{\rm i},t\right)\frac{dr_{\rm i}}{dt} ,
 \label{Eq: lm}
\end{equation}
where $A_{\rm g}$ is the effectiveness of gamma-ray trapping, and $Q = 1.6\times 10^{13}\left(Z/A\right)Z^{4/3}$ is the recombination energy per unit mass. $A_{\rm g}$ is significant in modeling the light curves of Type IIb and Ib/c SNe \citep{Nagy2014}. The optical depth of gamma-rays can be defined as $\tau_{\rm g} = A_{\rm g}/t^2$ \citep{Nagy2014, Chatzopoulos2012}.

\subsection{Assumptions made by \cite{1989ApJ...340..396A}} 
Two assumptions were made in  \cite{1989ApJ...340..396A} to separate the variables. 
For the spatial of the temperature profile, they assumed shape invariance of $\psi(x)$
\begin{equation}
	\int_{0}^{x_{\rm i}} \psi(x)x^2 dx = x_{\rm i}^3 \int_{0}^{1} \psi(x)x^2 dx = x_{\rm i}^3I_{\rm Th}^0, \label{Eq: approx2}
\end{equation}
and
\begin{equation}
	\frac{d\psi(x)}{dx}\bigg|_{x = x_{\rm i}}=-\frac{1}{x_{\rm i}}. \label{Eq: approx1}
\end{equation}
In most cases, it is not true to simply assuming $\psi(x)\propto \ln{\frac{1}{x}}$ as in Eq.~(\ref{Eq: approx1}). 
Letting the recombination front to be the pseudo-photosphere \citep{1989ApJ...340..396A, Nagy2014}, then Eq.~(\ref{Eq: approx2}) gives $I_{\rm ph} = x_{\rm ph}^3 I_{\rm Th}^0$. Taking the uniform density case as an example, we have $\psi(x)=sin(\pi x)/\pi x$ in this case. $I_{\rm ph}\approx x_{\rm ph}^3 I_{\rm Th}^0$ is valid only when $x$ tends to zero. In other cases, such approximation may lead to an incorrect result. {In our model, we abandon these approximations and calculate the L.H.S of Eq.~(\ref{Eq: approx2}) and (\ref{Eq: approx1}) numerically to avoid this problem.}

\subsection{The receding boundary condition}
In the literature, fixed boundary condition is assumed {(fast recombination approximation)}. As discussed in \cite{1980ApJ...237..541A}, it is only valid for dense objects. When the envelope expands, the density of the outer layer decreases rapidly. Then such an assumption might not be appropriate. If the relaxation time of the radial temperature profile of the envelope is short enough, Eddington's boundary condition is needed:
\begin{equation}
	acT^4=\frac{3}{4}T_e^4(\tau+q).
\end{equation} 
If we take $q = 1$, the Eddington approximation can 
be written as
\begin{equation}
	\frac{\partial \psi}{\partial x} = -\frac{3}{4}\psi_e\frac{R(t)}{\lambda(x)}.
\end{equation}
So at the photosphere
\begin{equation}
	\psi(x_{\rm ph})=-\frac{4}{3}\frac{\lambda(x_{\rm ph})}{R(t)}\frac{\partial \psi(x)}{\partial x}\bigg|_{x = x_{\rm ph}}. \label{Eq: photosphere boundary}
\end{equation}
  The expression of $\psi(x)$ can be obtained by solving Eq.~(\ref{Eq: spatial part}). Unfortunately it is only analytically solvable when the density distribution is uniform, i.e., $\eta(x) = 1$. In this case $\psi(x)=sin(\sqrt{\alpha} x)/\sqrt{\alpha} x$. 
  If the density is an exponential distribution, i.e., $\eta(x)=e^{A x}$. 
  Eq.~(\ref{Eq: spatial part}) can only be solved numerically. The solution of Eq.~(\ref{Eq: spatial part}) contains an unknown parameter $\alpha$, which is determined by substituting $\psi(x)$ into Eq.~(\ref{Eq: photosphere boundary}).

\cite{1993ApJ...414..712P} made a rough assumption: $\alpha = (\pi/x_{\rm i})^2$, which is obtained by assuming $\psi(x_{\rm i})=0$. As discussed above, $\psi(x_{\rm i})$ can not be zero. If so, the temperature at the recombination front is zero rather than $T_{\rm ion}$. However, the result under this assumption is similar to ours. In our model, $\alpha$ is roughly proportional to $x_{\rm ph}^{-2}$ and $x_{\rm ph}$  decreases with time. Thus  the results of \cite{1993ApJ...414..712P} implies the validity of our assumption.

Note that $\alpha$ is time-dependent in our model, which makes $\psi(x)$ also time-dependent. So Eq.~(\ref{Eq: spatial part}) should be solved at every time step, when numerical method is implemented. 

\section{Medium recombination approximation} \label{Sec: 3}
Because of the recession of the photosphere, $\alpha$ and $\psi(x)$ varies by the location of photosphere predicted by the Eddington surface boundary condition (\cite{2022ApJ...933....7J}). 
We need to determine how the temperature profile responds to the receding photosphere. Assuming the recombination wave as the photosphere, two approximations to the relaxation time of the temperature profile wave were studied by \cite{Arnett1996} (fast and slow approximations). If the recombination front moves slowly, photon diffusion inside the envelope will ensure the temperature profile adjusts to its new outer boundary with the same spatial profile \citep{1993ApJ...414..712P}. While for a fast-moving recombination font, the temperature profile inside will not react to the fact that the out part are being chopped off \citep{1989ApJ...340..396A, Nagy2014}.
Generally, the actual response of the temperature to the receding boundary is uncertain. {Because Eddington boundary condition may be inaccurate when envelope is optically thin \citep{Fukue2014}.} The real evolution of radial temperature profile should be between the slow and fast situations. 
Therefore, we define a parameter $\gamma$ to adjust the value of $\alpha$ in each time iteration { and fit with observation to obtain the value of $\gamma$}. The explicit form is written as
\begin{equation}
    \alpha^{i+1}_{(\rm cor)} = \alpha^i + \left(\alpha^{i+1}_{(\rm pre)} - \alpha^{i}\right)\gamma,
\end{equation}
where $\alpha^{i+1}_{(pre)}$ is calculated from Eq.~(\ref{Eq: photosphere boundary}). This is like the predictor-corrector methods in the numerical analysis. The parameter $\gamma$ is used to adjust the evolution of $\alpha$ to fit observations. Since explosion velocity $v_{\rm sc}$ can be obtained from observation. Substituting $v_{\rm sc}$ into our model and fitting the observed light curve, we obtain the value of $\gamma$. $\gamma > 1$ represents $\alpha$ evolves faster than the speed under the current time step. $1>\gamma>0$ represents $\alpha$ evolved slower than the current time step. $\gamma=0$ represents the constant $\alpha$ case, i.e., fast recombination approximation. 

The slow recombination approximation assumes the temperature profile evolves with photosphere recession without delay, while the fast one assumes a stationary spatial temperature distribution. They are two limit conditions corresponding to $\gamma = \infty$ and $\gamma = 0$. {Because $\gamma$ is finite, we name our model the medium recombination approximation.}

{We also find that without the constrain of $\gamma$, the evolution of $\alpha$ is related to the time steps in the numerical algorithm. There are two reasons. Firstly, the Eddington approximation may not be accurate at optically thin condition as \cite{Fukue2014} and \cite{2018ApJ...860...131C} suggested. 
Secondly, both the photosphere and the recombination front are related to the radial temperature profile, i.e. the value of $\alpha$. However, the value of $\alpha$ is not determined either. One more parameter is required to determine the evolution of $\alpha$. It can be time step \footnote{The time step must be small enough to avoid numerical error. In this paper, 0.25 days is usually short enough.} or the $\gamma$. The evolution of radial temperature profile is unique for any source. For any given time step, we are always able to adjust $\gamma$ to get same evolution of $\alpha(t)$, which makes $\alpha(t)$ independent of numerical method. \cite{1993ApJ...414..712P} used luminosity to determine the radius of recombination front. This method is originally proposed in \cite{imshennik_1992}. However, this way is so rough that can only reflect the scaling relation rather than strictly fit observation. One more parameter $t_{\rm i}$ is needed which determines the moment that recombination starts. In our model, time step and $\gamma$ also influence that moment, i.e., the value of $t_{\rm i}$. Therefore, they are similar way. But our method can fit observation better.}

{Considering that both the time-step of the numerical method and $\gamma$ effect the generated light curves, we fix the time step of all the runs in this paper to be $dt = 0.25\,\rm day$ except for the high resolution runs. In this way, we are able to compare the relative values of $\gamma$ for different sources. }

\begin{figure}[]
    \centering
    \includegraphics[width=\linewidth]{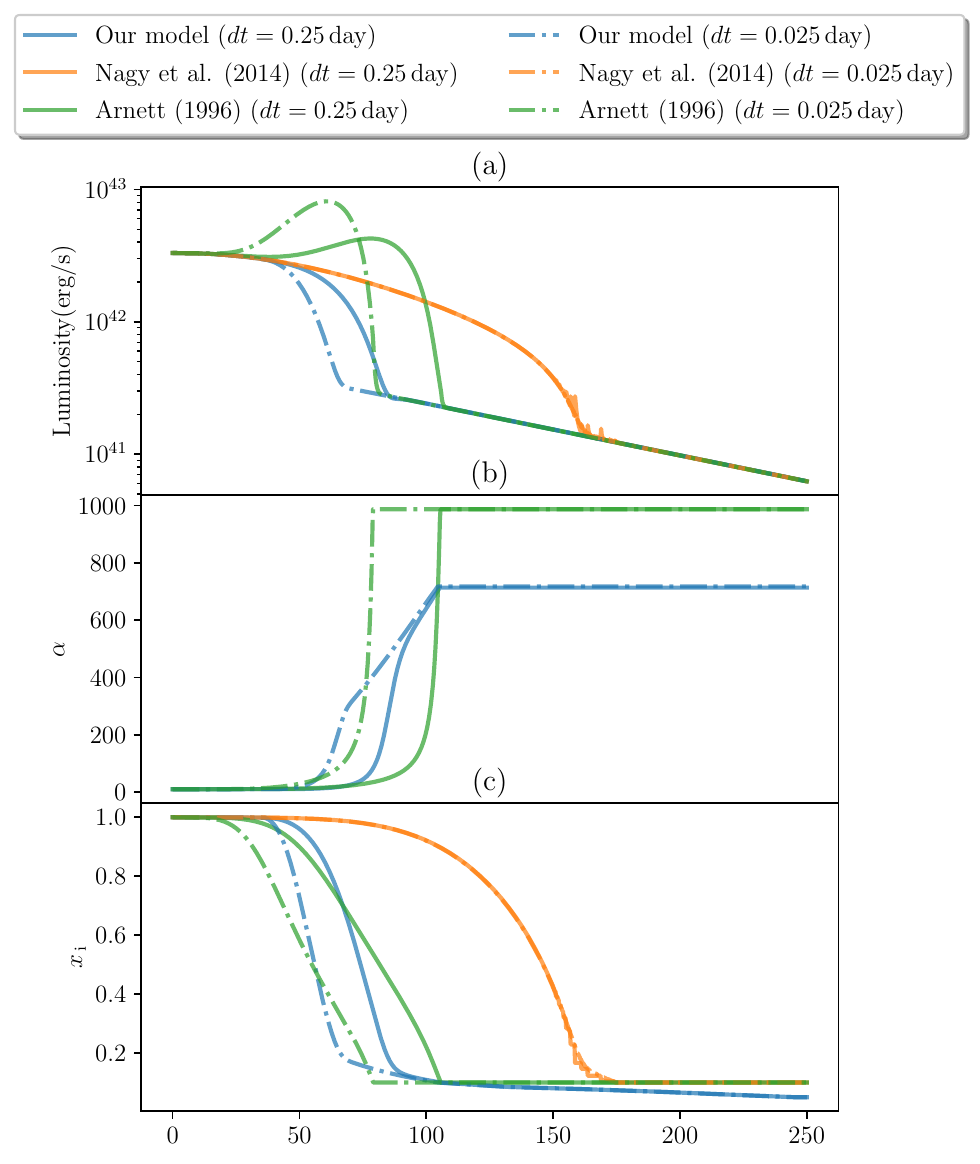}
\caption{Time evolution of luminosity and the parameters of $\alpha$ and $x_i$ in three models. {The blue lines are generated our model; the green lines are from the slow recombination model in \cite{Arnett1996}; the orange lines are generated from the model of \cite{Nagy2014}. } We adopt $\gamma=1$ in our model.}
\label{Fig: light curve compare.}
\end{figure}

\section{Model comparison with previous works} \label{Sec: 4}
\subsection{Comparison with \cite{Arnett1996} and \cite{Nagy2014}}
In Fig.1, we use the same parameters of $R_0=5\times10^{13}\,\rm cm$, $E_{\rm Th}^0 = E_{\rm kin} = 0.955\times10^{51}\,\rm  erg$, $\kappa_{\rm t} = 0.3 \,\rm cm^2/g$, $T_{\rm ion}=6000\,\rm K$, $M_{\rm Ni}=0.04\,\rm  M_{\odot}$, $M_{\rm ej} = 10\, \rm M_{\odot}$, $A = 0$, {$\gamma=1$ and the numerical time step of $dt=0.25\,\rm day$ }to compare our results with that of \cite{Arnett1996} and 
\cite{Nagy2014}. We exclude the recombination heat in the three models. We use the slow approximation with 
$\psi(x) = \sin(\pi x / x_{\rm i})/(\pi x / x_{\rm i})$ for the model of \cite{Arnett1996}.
In Fig.~\ref{Fig: light curve compare.}, we show the results from the three models with two different time steps of {0.25 day and 0.025 day}. We observe that $\alpha$ increases significantly during the recession of the photosphere, which also leads to significant change of $\psi(x)$ and $\phi(t)$ \citep{2022ApJ...933....7J}. The evolution of photosphere is crucial to determine the luminosity. 

There is a typo in the equation of A41 in \cite{1989ApJ...340..396A}, which is corrected in \cite{Arnett1996}. \cite{Nagy2014} still adopted the original version, which is written as:
\begin{equation}
	\frac{d\phi(t)}{dt} = \frac{R(t)}{R_{\rm 0} x_{\rm i}^3}\left[p_{\rm 1} \zeta(t)-x_{\rm i}\frac{\phi(t)}{\tau_{\rm d}}-2x_{\rm i}^2 \phi(t)\frac{R_{\rm 0}}{R(t)}\frac{dx_{i}}{dt}\right], \label{Eq: phi_Nagy}
\end{equation}
where $p_{\rm 1}$ is defined in Eq.~(\ref {Eq: phi}). From the assumption of Eq.~(\ref{Eq: approx2}) made by \cite{1989ApJ...340..396A}, the coefficient of the third term in the square bracket of Eq.~(\ref{Eq: phi_Nagy}) should be 3, i.e., $3x_{\rm i}^2 \phi(t)\frac{R_{\rm 0}}{R(t)}\frac{dx_{i}}{dt}$. 

{The approximations of Eq.~(\ref{Eq: approx2}) and Eq.~(\ref{Eq: approx1}) on $\psi(x)$ generate an initial bump in luminosity as shown in the green lightcurves in Fig.~\ref{Fig: light curve compare.}(a). The lightcurves of Fig.~13.12, 13.13, and 13.14 in \cite{Arnett1996} also show similar bumps. The two assumptions about $\psi(x)$ (Eq.~(\ref{Eq: approx1}) and (\ref{Eq: approx2})) overestimate $dx_{\rm i}/dt$, which make the recombination proceed faster. \cite{Arnett1996} did not consider the evolution of $\tau_{\rm d}$ but simply used a constant $\tau_{\rm d}$ (see Eq.~13.39 in \cite{Arnett1996}), which resulted in a longer plateau phase than our model. }

\begin{figure}[h]
    \centering
    \includegraphics[width=\linewidth]{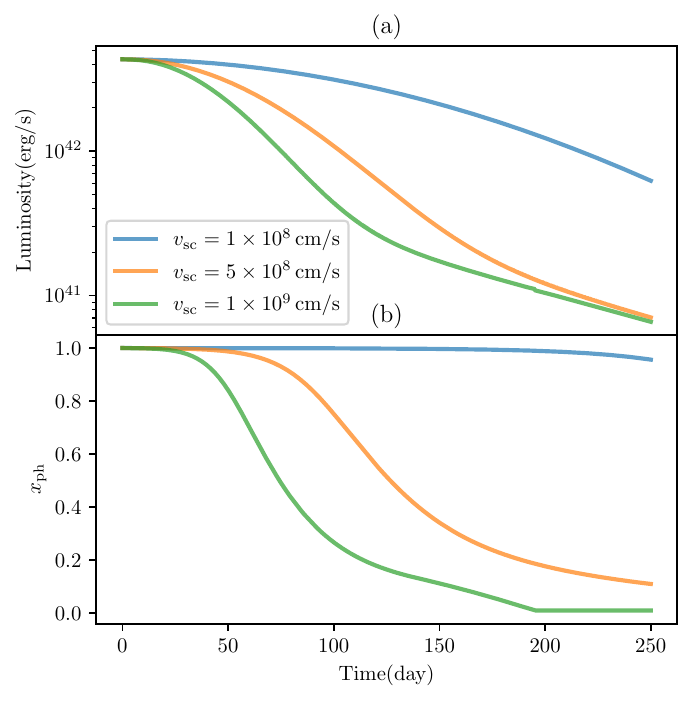}
 	\caption{Light curves (panel (a)) and evolution of the photosphere (panel (b)) of our model using the fast recombination approximation. The parameters are the same as in Fig.~\ref{Fig: light curve compare.}. Different colors indicates different explosion velocities. The blue, orange, and green lines represents the case with $v_{\rm sc}=1\times10^8$, $5\times10^8$, and $1\times10^9\,\rm cm/s$.} 
 	\label{Fig: no changing alpha}
 \end{figure}

{By defining $I_{\rm ph}$, we numerically calculate the integration and derivative of $\psi(x)$ without any approximations (see Eq.~(\ref{Eq: phi})). 
As long as the temperature profile evolves with the recession of the photosphere, we have time-variant $\alpha$. Along with the updated version of $d\phi(t)/dt$, we generate the light curves of typical SNe IIP. 
A natural question is, does Eq.~\ref{Eq: phi} still work under the fast approximation? 
We present the light curves calculated by our model with a constant $\alpha$ ($\gamma=0$) and different initial kinetic energies in Fig.~\ref{Fig: no changing alpha}, which are obviously different from SNe IIP. 
Therefore we conclude that the coevolution of $\alpha$ and boundary condition is essential for modeling the lightcurves of SNe IIP.}

\subsection{Comparison with \cite{1993ApJ...414..712P}}
\begin{figure}
    \centering
    \includegraphics[width=\linewidth]{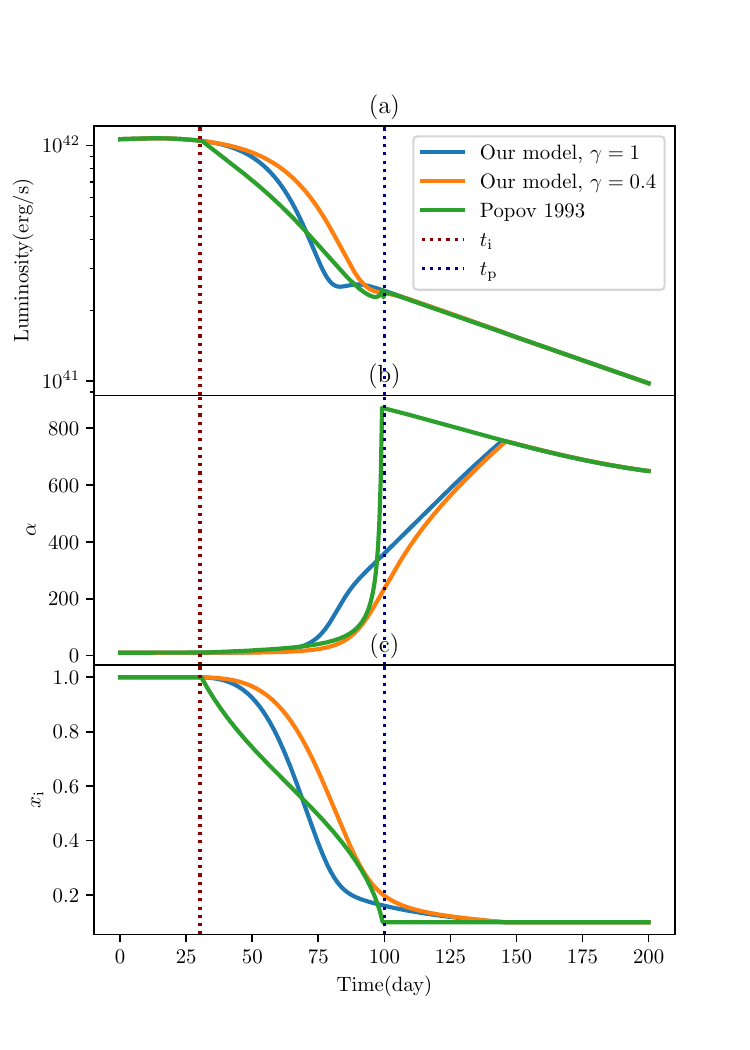}
\caption{Luminosity and parameter evolution of our model and \cite{1993ApJ...414..712P}. {The blue and orange lines come from our model with $\gamma=1$ and $0.4$ respectively; the green line is the result from \cite{1993ApJ...414..712P}. The time steps of all the models are $0.25\,\rm day$. The recombination begins at $t_i$ and stops at $t_p$, which are marked with red and blue dotted lines in the figure.}}
 	\label{Fig: popov}
\end{figure}

\cite{1993ApJ...414..712P} proposed that the luminosity given by diffusion approximation equals to the black body radiation of the photosphere (See Eq.~(11) in their paper). In this way, the temperature profile parameter $\alpha$ does not influence the position of the recombination front, which makes the model not affected by the time step. However, this approach requires another input parameter $t_{\rm i}$ for calculation, which is hard to obtain accurately. The analytical expression of the position of pseudo-photosphere ($x_{\rm i}$) is expressed as
\begin{equation}
    x_{\rm i}^2(t) = \frac{t_{\rm i}}{t}\left(1+\frac{t_{\rm i}^2}{3t_{\rm a}^2}\right)-\frac{t^2}{3t_{\rm a}^2}.
\end{equation}
where $t_{\rm i}$ is the moment when the recombination begins and $t_{\rm a}$ is a character time related to the diffusion and expansion time. In Fig.~\ref{Fig: popov}\footnote{The calculation of light curves in this figure does not include recombination heat and gamma-ray leakage.}, we compare our results with \cite{1993ApJ...414..712P} by using their parameters of $E = 10^{51}\,\rm erg$, $R_0 = 500\,\rm R_{\odot}$, $M_{\rm ej} = 10 \,\rm M_{\odot}$, $\kappa = 0.34$, $t_{\rm i} \approx t_{\rm a}/\sqrt{\Lambda} = 30.2\,\rm day$. The recombination wave defined in \cite{1993ApJ...414..712P} propagates slower than our model, giving lower slope in the light curve during the transition phase. Two different $\gamma$ values are implemented in Fig.~\ref{Fig: popov}, $\gamma=0.4$ agrees with \cite{1993ApJ...414..712P} better.
\begin{figure}
    \centering
    \includegraphics[width=\linewidth]{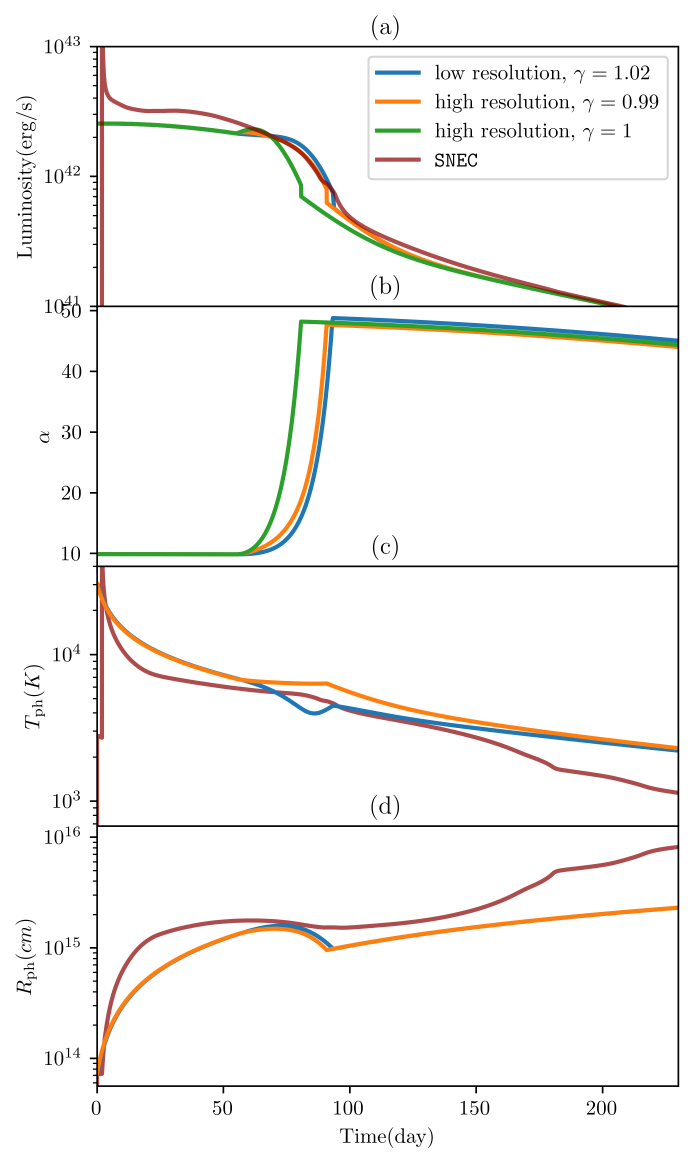}
    \caption{A comparison between the results from our semi-analytical model and that from the hydrodynamic simulations of {\tt SNEC}. {The time step for the low resolution run is $0.25\,\rm day$, and the high resolution run is $0.025\,\rm day$}.}     \label{fig: SNEC}
\end{figure}
{Although the propagation of recombination wave in \cite{1993ApJ...414..712P} is not influenced by numerical time step, a new parameter $t_{\rm i}$ is introduced in their work, which has similar effect as $\gamma$ in our model. Both of them controls the duration of the plateau phase. }

\subsection{Comparison with {\tt SNEC}}
{SNEC solves the hydrodynamics and diffusion radiation transport in the expanding envelopes of core-collapse supernovae, taking recombination effects and the decay of radioactive nickel into account. We compare our results with that of {\tt SNEC}, which is presented in Fig.~\ref{fig: SNEC}. Our model star is a red supergiant whose internal structure is derived from a $15\,\rm M_{\rm \odot}$ zero-age main sequence star obtained from the 1D stellar evolution code {\tt MESA} \citep{2013ApJS..208....4P}. The evolution of the model star was calculated by {\tt MESA}. The subsequent hydrodynamic evolutions were followed by SNEC. The {\tt SNEC} simulation is set up with a thermal bomb with an initial condition of: model mass $M=12.29\,\rm M_{\odot}$, initial radius $R_0=1038.71\,\rm R_{\odot}$, nickel mass $M_{\rm Ni}=0.04\,\rm M_{\odot}$, and total energy $E_{\rm tot}=10^{51}\,\rm erg$. The lower model mass in the initial condition for the explosion is due to the mass loss during {\tt MESA} simulation. {\tt SNEC} code calculates opacity in each grid point of the model from the existing Rosseland mean opacities for different components, temperature and densities of matter. The floor value for opacity needs to be determined. Using the default set up of {\tt SNEC} code, we set the opacity floor in the envelope (solar metallicity $Z_{\odot}=0.02$) to be $0.01\,\rm cm^2/g$ and core ($Z=1$) to be $0.24\,\rm cm^2/g$ \citep{2011ApJ...729...61B}, which is also adopted in \cite{Nagy2016}. Following \cite{Nagy2016}, we set the opacity in our semi-analytical model to be $0.24\,\rm cm^2/g$.}

\begin{figure*}
    \centering
    \includegraphics[height=0.24\linewidth]{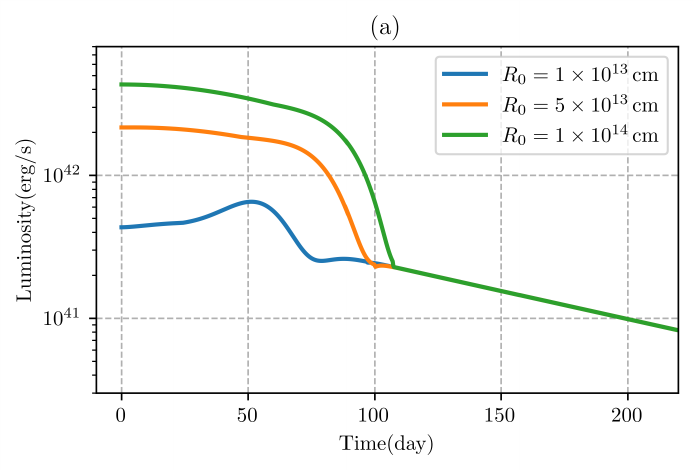}
    \includegraphics[height=0.24\linewidth]{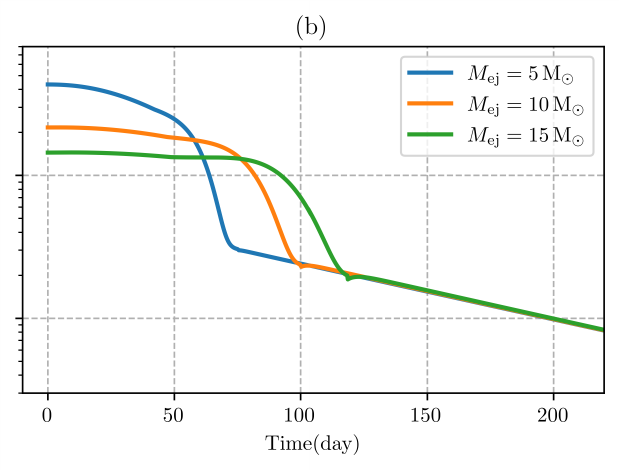}
    \includegraphics[height=0.24\linewidth]{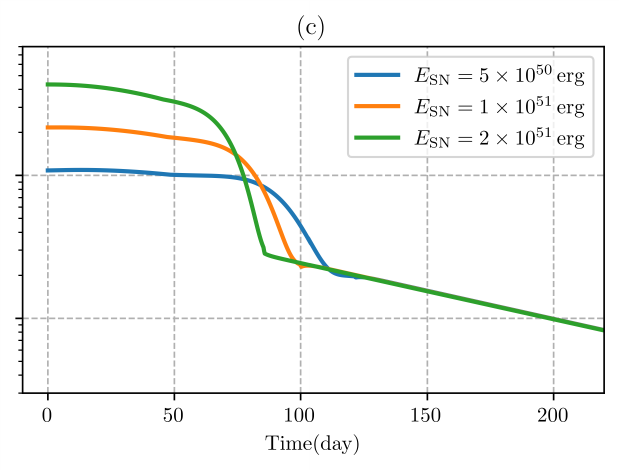}
    \includegraphics[height=0.24\linewidth]{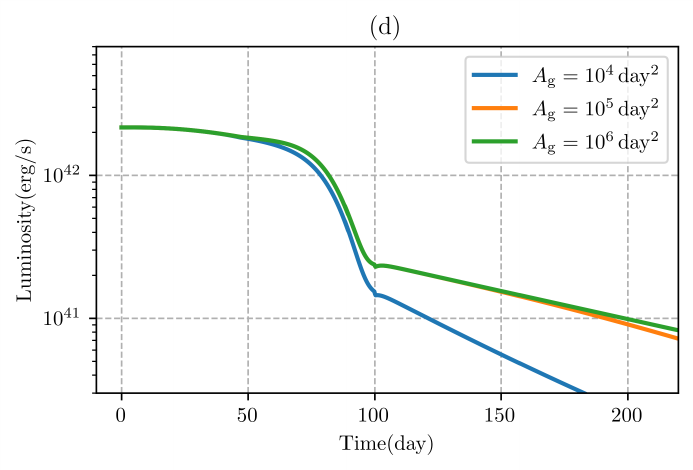}
    \includegraphics[height=0.24\linewidth]{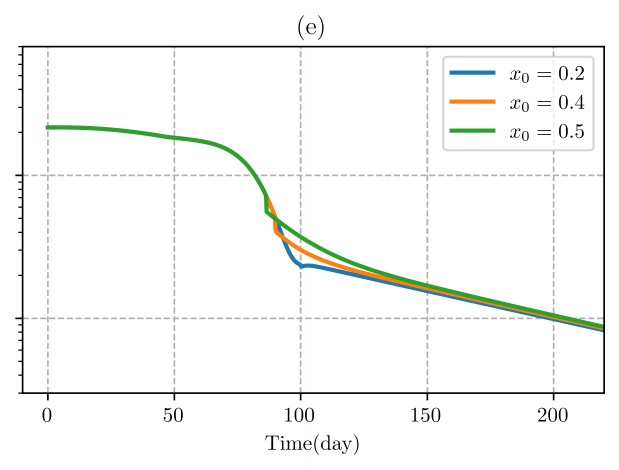}
    \includegraphics[height=0.24\linewidth]{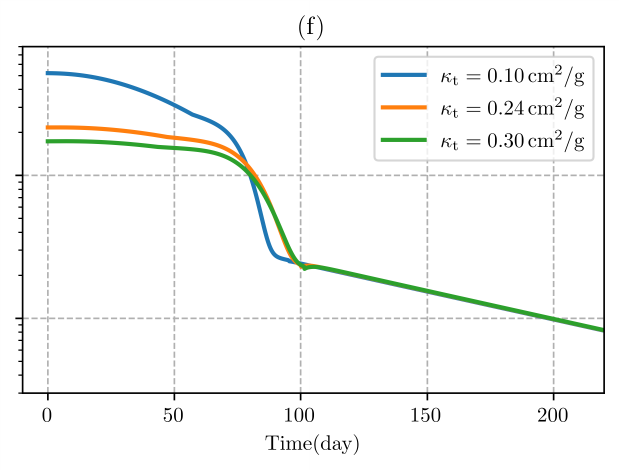}
    \includegraphics[height=0.24\linewidth]{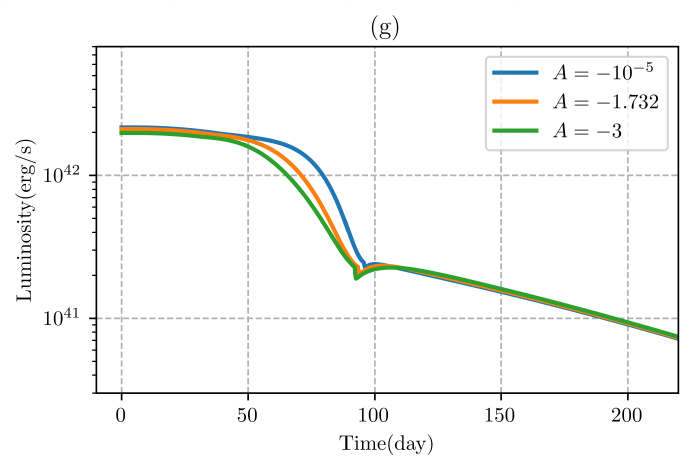}
    \includegraphics[height=0.24\linewidth]{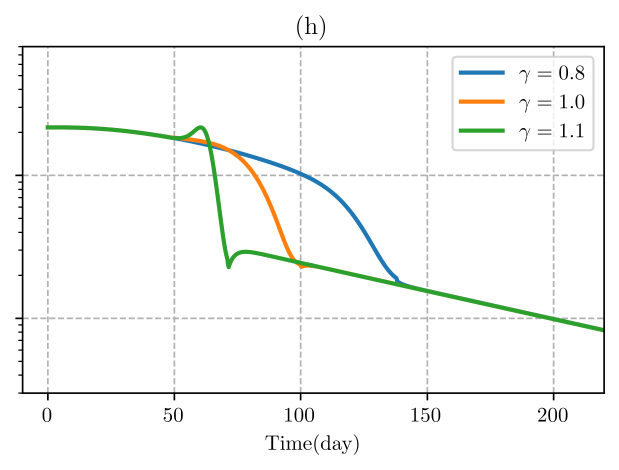}
    \caption{{In each panel, we investigate the impact of one single parameter on our model by varying its value while keeping all other parameters constant. The fiducial model is represented by the orange line in all panels, which shares the same parameter values as depicted in Fig.~\ref{Fig: light curve compare.}. The other two light curves show the effect of changing one parameter at a time. The investigated parameters are: (a) initial radius of envelope $R_0$; (b) ejecta mass $M_{\rm ej}$; (c) explosion energy $E_{\rm SN}$; (d) gamma-ray trapping effectiveness $A_{\rm g}$; (e) minimum recombination front $x_0$; (f) opacity $\kappa_{\rm t}$.; (g) exponential index of density distribution $A$; {(h) dimentionless index $\gamma$}.}}     \label{fig: initial parameters}
\end{figure*}

{For the initial condition of our numerical calculations, we adopt the same ejecta mass $M_{\rm ej}$, initial radius $R_0$, Nickel mass $M_{\rm Ni}$, and total energy $E_{\rm tot}$ as in {\tt SNEC} simulations. The time step used in Fig.~\ref{fig: SNEC} is $dt=0.25\,\rm day$ for low resolution run, and $dt=0.025\,\rm day$ for the high resolution run.}
As the blue and orange lines show in Fig.~\ref{fig: SNEC}, both the high and low resolution runs of our model matches the numerical result well. As discussed in Sec.~3, the evolution of $\alpha(t)$ determines the length of plateau phase, which is adjusted by $\gamma$. For the different time step runs in Fig.~\ref{fig: SNEC} (blue and orange lines), different values of $\gamma$ are chosen to match the numerical result. Although the values of $\gamma$ are different, the evolution of $\alpha(t)$ and the light curves are almost the same.

\section{The effects of different parameters} \label{Sec: 5}
In Fig.~\ref{fig: initial parameters}, light curves with different initial conditions are plotted. In each panel, the orange line represents our fiducial run which shares the same initial parameters as in Fig.~\ref{Fig: light curve compare.}. {The time step of all the models in this section is $dt=0.25\,\rm day$, as the same in previous sections.}

The light curves from SNe IIP with different ejecta masses are presented in Fig.~\ref{fig: initial parameters}(a). Although other parameters are the same, the SN ejecta with smaller initial radius has significantly lower luminosity during plaeatu phase. This result roughly agree with Stefan–Boltzmann law.

Fig.~\ref{fig: initial parameters}(b) represents the effect of different ejecta mass. Longer cooling time is required for the ejecta with higher mass, which leads to a longer plateau phase. {The denser envelope makes slower photosphere recession is also responsible for the longer plateau phase for the higher ejecta mass.} On the other hand, ejecta mass also has minor effect to the the slope during the transition phase. 

{Fig.~\ref{fig: initial parameters}(c) shows the effect of different explosion energies from the SNe under the assumption of equal kinetic energy and thermal energy. In general, lengths of the plateau phases of the SNe here are close, but luminosities are different. High explosion energy generates higher effective temperature and faster expanding. However, high expanding velocity causes high cooling rate which leads to shorter plateau phase. Thermal energy and kinetic energy compensate with each other. The difference of the plateau phase is hard to be too large. Main effect of the SN energy is the plateau luminosity.}

The effectiveness of gamma-ray trapping $A_{\rm g}$ mainly influence the slope of radioactive tail (see Fig.~\ref{fig: initial parameters}(d)). For the typical time scale of a supernova $t_{\rm scale}\sim 300\,\rm day$, when $A_{\rm g}$ is higher than $\sim10^5\,\rm day^2$, we are safe to assume all the gamma ray is trapped. For the effect of the radioactive decay of $\ce{^{56}Ni}$, our model shows similar result with \cite{Nagy2014}. Thus it is not presented in Fig.~\ref{fig: initial parameters}

{To avoid negative value for the recombination front $x_{\rm i}$, a minimum value $x_0$ for it is required. The physical meaning of it is that it corresponds to the boundary of the inner high dense and temperature core of the supernova. The effect of different $x_0$ is presented in Fig.~\ref{fig: initial parameters}(e). During the transition phase, it primarily affects the ending moment of the rapid decrease of luminosity. When $x_0\lesssim0.2$, the effect from this parameter is negligible.}

{In Fig.~\ref{fig: initial parameters}(f), the light curves from different opacity is plotted. Low opacity model generates higher plateau phase. The lengths of the plateau phases from different opacity are basically similar.}

{The effect of different density distributions is presented in Fig.~\ref{fig: initial parameters}(g). Different exponential index $A$ in $\eta(x)$ has relatively weak effect to the morphology of the light curve. The slope of the radial density distribution mainly affects the slope of the light curves during transition phase. }

The effect of $\gamma$ and explosion velocity ($E_{\rm kin}$), shows similar effect to the light curves. {As Fig.~\ref{fig: initial parameters}(h) shows, different $\gamma$ value makes different length of plateau phase, which is similar with the effect of explosion velocity (see Fig.~3a in \cite{Nagy2014}). The higher $\gamma$ or explosion velocity, the shorter generated plateau phase. The physics behind kinetic energy and $\gamma$ also has some similarity. The expansion of the envelope decreases the temperature inside it. While $\gamma$ implies the cooling rate of the envelope, higher $\gamma$ also leads to lower temperature. That is why both $E_{\rm kin}$ and $\gamma$ influence the length of the plateau.} Moreover, initial thermal energy $E_{\rm Th}^0$ \citep{Nagy2014} and initial radius $R_0$ also have almost the same effect to the light curve. {The degeneracy of the parameters has been found in \cite{Goldberg2019}.} To accurately determine the exact value of these parameters when fitting to a specific source, we manually input the explosion velocity $v_{\rm sc}$ into the model, and assume initial thermal energy equals kinetic energy ($E_{\rm Th}^0=E_{\rm kin}$). Define the energy of the supernova is $E_{\rm tot} = E_{\rm kin}+E_{\rm Th}^0$, {in which gravitational energy is neglected}. We have 6 parameters to determine the light curves, which are total energy $E_{\rm tot}$, ejecta mass $M_{\rm ej}$, mass of $\ce{^{56}Ni}$ $M_{\rm Ni}$, initial radius $R_0$, dimensionless parameter $\gamma$ and density distribution parameter $A$ if necessary. For most cases, we only use $M_{\rm Ni}$, $M_{\rm ej}$, $\gamma$ and $R_0$ as free parameters to fit, while total energy $E_{\rm tot}$ is partially given by observation (Only $M_{\rm ej}$ is unknown). 

\section{Two-component model} \label{Sec: 6}
A two-component model is proposed in \cite{Nagy2016} to explain the initial spike in the light curves of type-IIP supernovae. The density profile of the envelope is assumed to follow the broken power law, which has been widely adopted \citep{1982ApJ...258..790C, 1999ApJ...510..379M,2010ApJ...717..245K, 10.1093/mnras/stt1392}. In previous sections, we are focusing on the core part of the envelope. As suggested in \cite{Nagy2016}, the sparse shell component is essential for the initial spike at very early epoch. We adopt a broken power law density profile for the shell component as follows:
\begin{equation}
	\eta(x)=\left\{
\begin{aligned}
& \left(x/x_0\right)^{-\delta} \, 0\leq x\leq x_0, \\
& \left(x/x_0\right)^{-n} \, x_0\leq x\leq 1,
\end{aligned}
\right.
\end{equation}
where $\delta$ is usually assumed to be zero \citep{2018ApJ...868L..24L, Nagy2016}, $n$ is a free parameter to fit. For $n \sim 10$ is the SN Ib/Ic and SN Ia progenitors \citep{2018ApJ...868L..24L, 1999ApJ...510..379M,2010ApJ...717..245K, 10.1093/mnras/stt1392}. The photosphere $x_{\rm ph}$ in this case see \cite{2022ApJ...933....7J}. For a normal type IIP supernova, its light curve has three main components, i.e., the initial spike, plateau, and radioactive decay tail. The latter two are modeled with exponential/uniform density profiles in previous sections. For the initial spike, the light curve of a sparse shell is needed. The total luminosity of the two components produces the light curves of type IIP supernovae.

As shown in our previous work \citep{2022ApJ...933....7J}, the light curve of the broken power law density profile with index $n=10$ has a minimal temperature gradient, which makes the effect of photosphere recession negligible. This work considers both the photosphere recession and recombination into both components of the model. In the core region, the recombination front and photosphere are very close, which makes the recombination front can be treated as pseudo-photosphere. However, in the shell region, the photosphere recedes deep in the shell ($x_{\rm ph}\sim 0.2$) before recombination starts. It's incorrect to set the recombination front as photosphere in this case. If so, before recombination starts (within 15 days after explosion), the recombination front is the edge of the ejecta, whose speed is precisely the speed of the edge of the envelope $v_{\rm sc}$. While the envelope is expanding, there is slight deceleration, which makes $v_{\rm sc}$ can be approximated as a constant. Since observation in \cite{2022MNRAS.513.4556Z} shows a rapid decrease of photospheric velocity. Thus setting recombination front to be pseudo-photosphere is not appropriate in this case.

\section{Fitting with observation} \label{Sec: 7}
{In this section, we present our model fitting of five sources and comparison of the best fit parameters between our model and former works. For all the runs we assume $\kappa=0.3\,\rm cm^2/s$ for the core component, and time steps $dt$ are all $0.25\,\rm day$. }
\subsection{SN~2016gfy}

\begin{figure}[h]
    \centering
    \includegraphics[width=\linewidth]{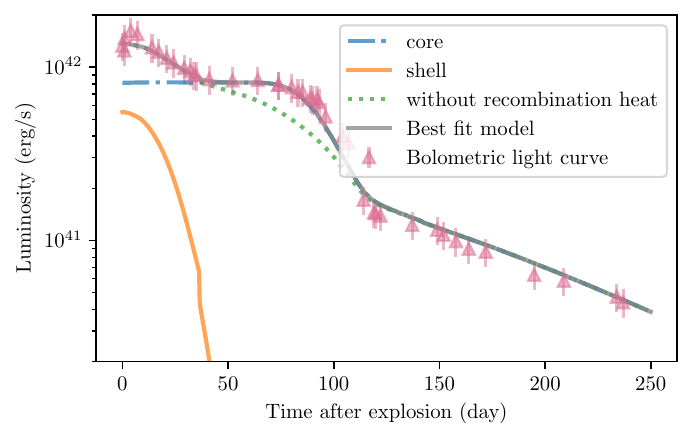}
 	\caption{{The fitting result of the bolometric light curve of SN~2016gfy. The blue dashed dotted line represents the contribution from core component of our model; orange solid line is the shell component; the green dotted line is the photospheric luminosity without the contribution of recombination heat; the grey solid line is the total luminosity from our model, and the pink error bars are the bolometric data from \cite{Singh2019}.}}
 	\label{Fig: SN2016gfy}
 \end{figure}

\begin{table}[h]
\begin{tabular}{llll}
\hline
Parameter                            & \multicolumn{2}{l}{Our model} & Literature$^*$   \\
                                     & core          & shell         &                  \\ \hline
$\gamma$                             & 0.77          & -             & -                \\
$M_{\rm ej} (\rm M_{\odot})$         & 13.6          & 0.5           & 14.6             \\
$E_{\rm tot} (\rm 10^{51}\,\rm erg)$ & 2.86          & -             & -                \\
$E_{\rm k} (\rm 10^{51}\,\rm erg)$   & 1.42          & 0.05          & 0.9-1.4          \\
$E_{\rm th} (\rm 10^{51}\,\rm erg)$  & 1.42          & 0.001         & -                \\
$M_{\rm Ni} (\rm M_{\odot})$         & 0.031         & -             & $0.033\pm 0.003$ \\
$R_{\rm 0} ( \rm R_{\odot} )$        & 160           & 503           & $\sim 350-700$  \\
$n$                                  & -             & 4             & -                \\ \hline
\end{tabular}
\caption{Fitting result for SN~2016gfy. $^*$ The estimated parameters come from \cite{Singh2019}.}
\label{table: table_2016gfy}
\end{table}

{We present our light curve fitting with the bolometric luminosity of SN~2016gfy in Fig.~\ref{Fig: SN2016gfy}, and fitting parameters are in Table~\ref{table: table_2016gfy}.}
{The bolometric data from SN 2016gfy comes from \cite{Singh2019}. The explosion velocity of the core component is $4200\,\rm km/s$, which is chosen from the speed of Fe II line at the end of plateau phase. Although the effect of Nickel mixing is considered important in \cite{Singh2019}, our model is able to fit the bolometric light curve well without adding this effect. Only from the morphology of the light curve may not enough to demonstrate the importance of Nickel mixing, which requires more detailed study \citep{Nakar2016}. We point out here, the reason why the {\tt LC2} \citep{Nagy2014} does not fit SN 2016gfy well may come from the previously discussed assumptions about $\psi(x)$ from \cite{1989ApJ...340..396A} and lacking the effect of photosphere recession \citep{2022ApJ...933....7J}.}

{The physical reason that the our model provides relatively flat plateau phase rather than slowly decreasing, comes from the contribution of recombination heat (the second term in Eq.~(\ref{Eq: lm})). Since recombination happens outside of photosphere (see Eq.~(\ref{Eq: optical depth}) and \cite{2018ApJ...868L..24L}). The energy from the Hydrogen recombination is directly added to the total luminosity \citep{Nagy2014}.  The red dotted line in Fig.~\ref{Fig: SN2016gfy} shows the light curve without the recombination heat, which demonstrates the flattened plateau is contributed by recombination heat.}

\subsection{SN~2019va}

\begin{figure}[h]
    \centering
    \includegraphics[width=\linewidth]{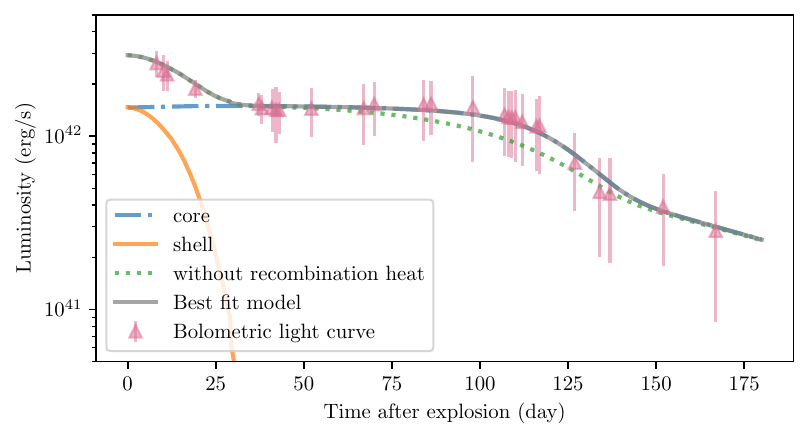}
 	\caption{{The fitting result for SN~2019va. The lines and dots share the same meaning with that of Fig.~\ref{Fig: SN2016gfy}.}} 
 	\label{Fig: SN2019va}
 \end{figure}
\begin{table}[h]
\centering
\begin{tabular}{llll}
\hline
Parameter                            & \multicolumn{2}{l}{Our model} & Literature$^*$   \\
                                     & core          & shell         &                  \\ \hline
$\gamma$                             & 0.87          & -             & -                \\
$M_{\rm ej} (\rm M_{\odot})$         & 19.59         & 0.3           & 17.1             \\
$E_{\rm tot} (\rm 10^{51}\,\rm erg)$ & 2.54          & -             & -                \\
$E_{\rm k} (\rm 10^{51}\,\rm erg)$   & 1.27          & 0.04          & 1.15             \\
$E_{\rm th} (\rm 10^{51}\,\rm erg)$  & 1.27          & 0.001         & -                \\
$M_{\rm Ni} (\rm M_{\odot})$         & 0.098         & -             & $0.088\pm 0.018$ \\
$R_{\rm 0} ( \rm R_{\odot} )$        & 466.56        & 1006.18       & 289              \\
$n$                                  & -             & 4             & -                \\ \hline
\end{tabular}
\caption{Fitting result for SN~2019va. $^*$ The estimated parameters come from \cite{2022MNRAS.513.4556Z}.}
\label{table: table_2019va}
\end{table}
The fitting result of SN~2019va is presented in Fig.~\ref{Fig: SN2019va}.
The bolometric data of SN 2019va comes from \cite{2022MNRAS.513.4556Z}. From their work the explosion velocity is $\sim 3000\,\rm km/s$. We choose $v_{\rm sc}=3300\,\rm km/s$ for the calculation of light curve in Fig.~\ref{Fig: SN2019va}. {The fitting parameters and comparison with the value provided in \cite{2022MNRAS.513.4556Z} are presented in Table.~\ref{table: table_2019va}. The shell component in our result has much larger initial radius than the value provided in \cite{2022MNRAS.513.4556Z}. The parameters estimated in \cite{2022MNRAS.513.4556Z} comes from the empirical relation \citep{1985SvAL...11..145L} that uses the data at $50\,\rm day$ after the explosion, which can not reflect the information when the outer shell is dominant (less than $30\,\rm day$ after the explosion). Therefore, their result only reflects the size of the core component.}

\subsection{SN~1999em}
\begin{figure}[h]
    \centering
    \includegraphics[width=\linewidth]{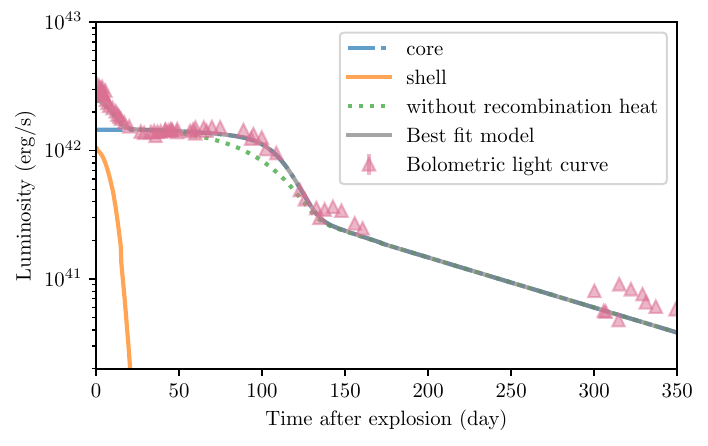}
 	\caption{{The fitting result for SN~1999em. The lines and dots share the same meaning with that of Fig.~\ref{Fig: SN2016gfy}.}} 
 	\label{Fig: SN1999em}
 \end{figure}
\begin{table}[h]
\centering
\begin{tabular}{lllll}
\hline
Parameter                            & \multicolumn{2}{l}{Our model} & Literature$^1$              & Literature$^2$             \\
                                     & core          & shell         &                             &                            \\ \hline
$\gamma$                             & 0.85          & -             & -                           & -                          \\
$M_{\rm ej} (\rm M_{\odot})$         & 19            & 0.2           & $19.10^{-6.18}_{+3.57}$     & $13.16^{-2.04}_{+4.94}$    \\
$E_{\rm tot} (\rm 10^{51}\,\rm erg)$ & 3.27          & -             & -                           & -                          \\
$E_{\rm k} (\rm 10^{51}\,\rm erg)$   & 1.64          & 0.01          & $4.21^{-1.55}_{+4.10}$      & $2.52^{-0.78}_{+2.19}$     \\
$E_{\rm th} (\rm 10^{51}\,\rm erg)$  & 1.64          & 0.06          & $0.32^{-0.05}_{+9.98}$      & $3.47^{-3.21}_{+4.85}$     \\
$M_{\rm Ni} (\rm M_{\odot})$         & 0.06          & -             & $0.06$                      & $0.04$                     \\
$R_{\rm 0} ( \rm R_{\odot} )$        & 350           & 500           & $947.25^{-918.50}_{+53.18}$ & $74.74^{-46.00}_{+927.12}$ \\
$n$                                  & -             & 6             & -                           & -                          \\ \hline
\end{tabular}
\caption{Fitting result for SN~1999em. Column $^1$ and column $^2$ are the two different sets of fitting result in \cite{2020MNRAS.496.3725J}.}
\label{table: table_1999em}
\end{table}

{
The fitting result for SN~1999em is presented in Fig.~\ref{Fig: SN1999em}. The original multiband data comes from \cite{Leonard_2002, Leonard_2003, 10.1046/j.1365-8711.2003.06150.x}. The bolometric data is from \cite{2022MNRAS.513.4556Z}, which is obtained using the {\tt Superbol} package \citep{Nicholl_2018}. For the light curve fitting, we choose the explosion velocity $v_{\rm sc}=3800\,\rm km/s$, which is the photospheric velocity $50\,\rm day$ after the explosion \citep{10.1046/j.1365-8711.2003.06150.x}. The fitting parameter and comparison with other literature are listed in Fig.~\ref{table: table_1999em}.  In general, our model provide similar parameters with the two different sets in \cite{2020MNRAS.496.3725J}. The parameters in \cite{2020MNRAS.496.3725J} shows large uncertainty about the initial radius of the ejecta. Our result is between the two sets and within the uncertainty.
}

\subsection{SN~2004et}

\begin{figure}[h]
    \centering
    \includegraphics[width=\linewidth]{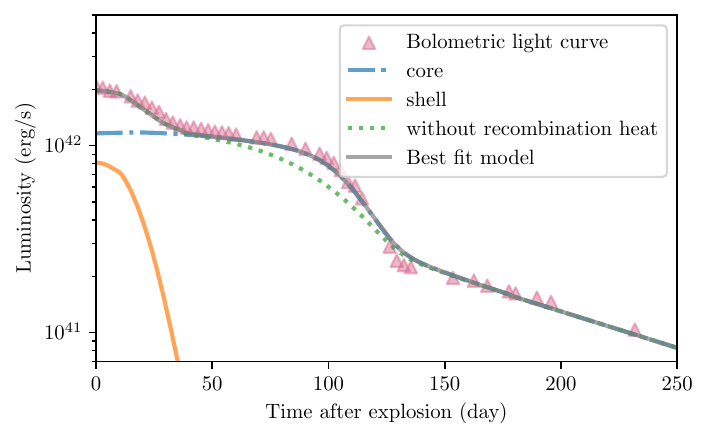}
 	\caption{{The fitting result for SN~2004et. The lines and dots share the same meaning with that of Fig.~\ref{Fig: SN2016gfy}.}} 
 	\label{Fig: SN2004et}
 \end{figure}
\begin{table}[h]
\centering
\begin{tabular}{lllll}
\hline
Parameter                            & \multicolumn{2}{l}{Our model} & Literature$^1$ & Literature$^2$ \\
                                     & core          & shell         &                &                \\ \hline
$\gamma$                             & 0.8           & -             & -              & -              \\
$M_{\rm ej} (\rm M_{\odot})$         & 12.0          & 0.8           & 11.0           & 14.0           \\
$E_{\rm tot} (\rm 10^{51}\,\rm erg)$ & 1.75          & -             & 1.95           & 0.88           \\
$E_{\rm k} (\rm 10^{51}\,\rm erg)$   & 0.88          & 0.0006        & -              & -              \\
$E_{\rm th} (\rm 10^{51}\,\rm erg)$  & 0.88          & 0.07          & -              & -              \\
$M_{\rm Ni} (\rm M_{\odot})$         & 0.053         & -             & $0.06$         & $0.036$        \\
$R_{\rm 0} ( \rm R_{\odot} )$        & 330           & 480           & 603.71         & 631.02         \\
$n$                                  & -             & 3             & -              & -              \\ \hline
\end{tabular}
\caption{Fitting result for SN~2004et. The fitting parameters from column $^1$ are from \cite{Nagy2014} and column $^2$ are from \cite{2010MNRAS.404..981M}.}
\label{table: table_2004et}
\end{table}
{
The light curve fitting for SN~2004et is presented in Fig.~\ref{Fig: SN2004et}, and fitting parameters and comparison with literature \citep{Nagy2014,2010MNRAS.404..981M} are listed in Table.~\ref{table: table_2004et}. We choose the explosive velocity to be $3500\,\rm km/s$ which is the mid-plateau velocity $\sim 3560\,\rm km/s$ in \cite{2006MNRAS.372.1315S}. The bolometric data is obtained from \citep{Nagy2016}. We observe a longer shell component compared to other sources in this paper, which accounts for the higher shell mass in our model.}

\subsection{SN~2005cs}

\begin{figure}[h]
    \centering
    \includegraphics[width=\linewidth]{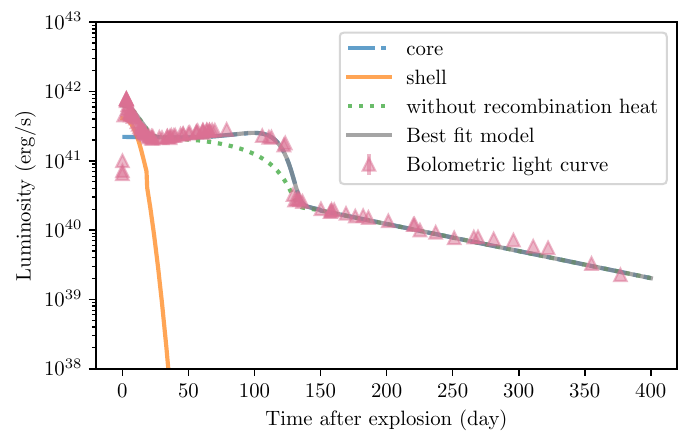}
 	\caption{{The fitting result for SN~2005cs. The lines and dots share the same meaning with that of Fig.~\ref{Fig: SN2016gfy}.}} 
 	\label{Fig: SN2005cs}
 \end{figure}
\begin{table}[h]
\centering
\begin{tabular}{llll}
\hline
Parameter                            & \multicolumn{2}{l}{Our model} & Literature$^*$ \\
                                     & core          & shell         &                \\ \hline
$\gamma$                             & 0.925          & -             & -              \\
$M_{\rm ej} (\rm M_{\odot})$         & 8.7           & 0.3           & 8.61           \\
$E_{\rm tot} (\rm 10^{51}\,\rm erg)$ & 0.32          & -             & 0.3            \\
$E_{\rm k} (\rm 10^{51}\,\rm erg)$   & 0.20          & 0.025         & -              \\
$E_{\rm th} (\rm 10^{51}\,\rm erg)$  & 0.12          & 0.0025        & -              \\
$M_{\rm Ni} (\rm M_{\odot})$         & 0.005         & -             & 0.0018         \\
$R_{\rm 0} ( \rm R_{\odot} )$        & 201.24        & 450           & 175.36         \\
$n$                                  & -             & 4             & -              \\ \hline
\end{tabular}
\caption{Fitting result for SN~2005cs. $^*$ The estimated parameters come from \cite{2006A&A...460..769T}.}
\label{table: table_2005cs}
\end{table}

{
We present the light curve fitting in Fig.~\ref{Fig: SN2005cs}. The fitting parameters as well as comparison with \cite{2006A&A...460..769T} are listed in Table~\ref{table: table_2005cs}. The original multiband data comes from \cite{10.1111/j.1365-2966.2006.10587.x, 10.1111/j.1365-2966.2009.14505.x}. Similar with SN~1999em, the bolometric data is calculated with {\tt Superbol} code \citep{Nicholl_2018} and obtained from \cite{2022MNRAS.513.4556Z}. For this source, the lighten up at the end of the plateau is relatively larger than other sources. The contribution of recombination heat is higher (the area between the black solid line and red dotted line). The relatively larger $M_{\rm Ni}$ in our model compared with \cite{2006A&A...460..769T} is from the unusually low \ce{^{56}Ni} in this source, which makes it very sensitive to the luminosity during radioactive tail.}

\section{Discussion and conclution} \label{Sec: 8}
{We integrate a receding photosphere into the previous homologous expansion model and derive the luminosity in a self-consistent way by introducing a medium recombination approximation.
We find that the photosphere is strongly related to the recombination front. The recession of the recombination front causes a significant recession of the photosphere. Using the Eddington boundary condition on the photosphere, we find the eigenvalue $\alpha$ is roughly proportional  to $x_{\rm ph}^{-2}$ as previous works suggested. At the end of the recombination, $\alpha$ is significantly larger than its initial value. Our medium recombination approximation allows us to determine the model parameters according to the observational data.  Moreover, our result is consistent with that of the widely used radiation-hydrodynamic simulations ({\tt SNEC}).}    

{It is also suggested that magnetars may power the explosions SNe IIP \cite{2017MNRAS.472..224S, 2018A&A...619A.145O}. In \cite{Nagy2014}, the nature of light curves produced by Arnett's analytical model with magnetar powering is briefly investigated. In their work, to get the typical light curves of SNe IIP, the rotational energy of the neutron star must be comparable to the recombination energy. Otherwise, there will not be a plateau in the light curves. This result is consistent with \cite{2017MNRAS.472..224S}, they suggest the strength of the magnetic field should be strong (greater than $10^{15}$ G), and the rotation period is several milliseconds. Using the relations of the spin-down time scale and the initial rotational energy of the magnetar, the corresponding scale and the rotational energy in  \cite{2017MNRAS.472..224S} is $\sim 0.4$ days and $\sim 5\times 10^{49} \rm erg$. Applying this result to our model, we found the influence from magnetar can hardly be distinguished. Therefore we do not present a figure with magnetar here. Since the energy injection of the magnetar in both our model and \cite{Nagy2014} are unnecessary. Whether or not SNe IIP are powered by magnetars remains unclear. It still needs more observation in the very early epoch.}

Our model can fit light curves of some SNe IIP  quite well. For some SNe IIP such as SN 2019va (which is shown in Fig.~\ref{Fig: SN2016gfy}, \ref{Fig: SN2019va} and \ref{Fig: SN2005cs}), their light curve increases mildly at the end of the plateau phase. \cite{2022MNRAS.513.4556Z} consider such effect comes from the distribution of \ce{^{56}Ni} in the envelope. However, the recombination heat also has a similar effect that lightens up the plateau as discussed in this paper. This puts an alternative channel to explain the flattened and longer plateau other than the Nickel mixing \citep{Kozyreva2019}. We will investigate it in detail in our future work.

\acknowledgments
We thank Dr. Jozsef Vinko and Dr. Andrea Nagy for providing us their source code. We also thank Dr. Xinghan Zhang for the useful discussion and bolometric data of SN 2019va, SN 1999em and SN 2005cs. We thank Dr. Singh for the bolometric data of SN 2016gfy.

\bibliographystyle{aasjournal}

\end{document}